\documentclass{aa}

\usepackage{graphicx}
\usepackage{natbib}
\usepackage{txfonts}
\bibpunct{(}{)}{;}{a}{}{,}    

\def\deg{$^{\rm o}$}

\begin{document}

\title{Orbital parameters of extrasolar planets derived from polarimetry}


\titlerunning{Orbital parameters of extrasolar planets derived from polarimetry}

\author{D.~M.~Fluri\inst{1} \and S.~V.~Berdyugina\inst{2,3}}

\authorrunning{D.~M.~Fluri and S.~V.~Berdyugina}

\institute{
Institute of Astronomy, ETH Zurich, CH-8093 Zurich, Switzerland 
\and 
Kiepenheuer Institut f\"ur Sonnenphysik, Sch\"oneckstr. 6, D-79104 Freiburg, Germany
\and 
  Tuorla Observatory, University of Turku, FIN-21500 Piikki\"o, Finland\\
  \email{fluri@astro.phys.ethz.ch, sveta@kis.uni-freiburg.de}
}


\date{Received $<$date$>$; accepted $<$date$>$}


\abstract
{Polarimetry of extrasolar planets becomes a new tool for their investigation, which requires the development of diagnostic techniques and 
parameter case studies.}
{Our goal is to develop a theoretical model which can be applied to interpret polarimetric observations of extrasolar planets. Here we present a theoretical parameter study that shows the influence of the various involved parameters on the polarization curves. Furthermore, we investigate the robustness of the fitting procedure. We focus on the diagnostics of orbital parameters and the estimation of the scattering radius of the planet.}
{We employ the physics of Rayleigh scattering to obtain polarization curves of an unresolved extrasolar planet. Calculations are made for two cases: (i) assuming an angular distribution for the intensity of the scattered light as from a Lambert sphere and for polarization as from a Rayleigh-type scatterer, and (ii) assuming that both the intensity and polarization of the scattered light are distributed according to the Rayleigh law. We show that the difference between these two cases is negligible for the shapes of the polarization curves. In addition, we take the size of the host star into account, which is relevant for hot Jupiters orbiting giant stars.}
{We discuss the influence of the inclination of the planetary orbit, the position angle of the ascending node, and the eccentricity on the linearly polarized light curves both in Stokes $Q/I$ and $U/I$. We also analyze errors that arise from the assumption of a point-like star in numerical modeling of polarization as compared to consistent calculations accounting for the finite size of the host star. We find that errors due to the point-like star approximation are reduced with the size of the orbit, but still amount to about 5\% for known hot Jupiters. Recovering orbital parameters from simulated data is shown to be very robust even for very noisy data because the polarization curves react sensitively to changes in the shape and orientation of the orbit.}
{The proposed model successfully diagnoses orbital parameters of extrasolar planets and can also be applied to predict polarization curves of known exoplanets. Polarization curves of extrasolar planets thus provide an ideal tool to determine parameters that are difficult to obtain with other methods, namely inclination and position angle of the ascending node of orbits as well as true masses of extrasolar planets.}


\maketitle


\section{Introduction}\label{sec:intro}
 
Direct observations of extrasolar planets succeeded only recently by measuring infrared thermal emission with the Spitzer Space Telescope \citep{charbonneauetal2005, demingetal2005, demingetal2006, harringtonetal2006}, polarimetric measurements \citep{berdyuginaetal2008a}, and direct imaging \citep{maroisetal2008}.
Polarimetry as a method to investigate extrasolar planets is particularly valuable because it involves direct observation of light scattered by the planetary atmosphere and reveals information about the chemical composition, the physical structure, and the evolution of the planetary atmosphere. So far the identification of atmospheric constituents is limited to transmission spectroscopy of transiting extrasolar planets \citep{charbonneauetal2002, vidalmadjaretal2003, vidalmadjaretal2004, tinettietal2007}, while polarimetry is capable of expanding these studies for non-transiting planets.

The first successful polarimetric detection of an extrasolar planet by  \citet{berdyuginaetal2008a} (see also new data and discussion on non-detections in \citeauthor{berdyuginaetal2009} \citeyear{berdyuginaetal2009}) opens new perspectives for diagnosing the physical properties of planetary atmospheres and underlines the urgent need for the development of theoretical tools to interpret the data.
Polarization signals of extrasolar planets have already been numerically simulated by several authors, who in particular investigated the influence of various scattering particles resulting from different atmospheric compositions. \citet{seageretal2000}, \citet{saarseager2003}, and \citet{houghlucas2003} have discussed close-in extrasolar giant planets with orbital radii smaller than 0.05~AU, which are thus unresolved and relatively hot, while \citet{stametal2004} studied the flux and polarized spectra of resolved Jupiter-like planets in larger orbits. In contrast to others \citet{senguptamaiti2006} have considered elliptical rather than circular orbits to calculate the polarization in the R passband due to scattering by water and silicate condensates, taking the planetary oblateness also into account. 
 
Most of these previous numerical simulations present light curves or spectra of the total degree of polarization only. They additionally are restricted by the approximation of the host stars as point-sources. However, the actual observables obtained from polarimeters are the Stokes parameters, not the total degree of polarization. It is obvious that part of the information concerning the orbital parameters is lost when forming the total degree of polarization. Moreover, the polarimetric method is well suited to detect extrasolar planets that closely orbit giant stars where the description of the host star as a point-source becomes questionable.

The numerical work has mainly concentrated so far on the red and near infrared parts of the spectrum, where the detailed composition of the atmosphere and the shape and size of the scattering particles play an important role. 
Also, the unsuccessful polarimetric detections of exoplanets were reported for red and green wavelengths \citep{lucasetal2009,wiktorowicz2009}.
The first successful detection of polarization from an extrasolar planet was obtained in the B and U passbands though \citep{berdyuginaetal2008a,berdyuginaetal2009} for HD189733b. As thoroughly discussed and modeled by \citet{berdyuginaetal2009}, this clearly indicates an atmosphere in which Rayleigh scattering plays a dominant role. The same conclusion was reached in a multi wavelength analysis of planetary transit curves \citep{pontetal2008,lecavelieretal2008}. 
In this atmosphere, a maximum polarization amplitude in green (red) wavelengths is at least a factor of two (ten) smaller than that in the B-band.
This would explain the non-detections by \citet{lucasetal2009} and \citet{wiktorowicz2009}. In this paper we concentrate on Rayleigh scattering and exclude Mie scattering, as the latter has not yet been detected in any extrasolar planet. We leave the case of Mie scattering for later studies. 

We discuss numerical simulations of polarized light curves obtained from Rayleigh scattering on unresolved extrasolar planets, particularly actually observed Stokes $Q/I$ and $U/I$ phase curves. The modeling strategy introduced here has been applied for the interpretation of the recently obtained polarimetric data for HD~189733b \citep{berdyuginaetal2008a}. We describe here the numerical model in detail, perform a theoretical parameter study for involved parameters, verify the validity of the Lambert sphere approximation, and test the stability of the procedure to fit observations. At this stage we concentrate on orbital parameters, in general for eccentric orbits, and on the planetary radius, which represents the first step when interpreting data and is sufficient for the currently available observations and polarimetric precision. Particular attention is paid to the correct treatment of the size of the host star.


\section{Modeling polarization of extrasolar planets}\label{sec:model}

\subsection{Model assumptions}\label{subsec:assumptions}

In this section we describe the numerical model for computing the polarization signal from extrasolar planets. We employ the following main assumptions:
\begin{itemize}
	\item the intensity of the scattered light is given either by the Lambert sphere or by Rayleigh scattering,
	\item the polarization is defined by the Rayleigh scattering law,
	\item the host star has a finite size,
	\item the direct irradiance from the star is unpolarized,
	\item the planet is unresolved from the host star,
	\item the radius of the planet is much smaller than the stellar radius,
	\item the thermal emission from the planet is neglected.
\end{itemize}

The first two assumptions form two cases: (i) a combination of the Lambert sphere angular distribution for intensity of the scattered light and the Rayleigh phase function for polarization and (ii) scattering consistently described by the Rayleigh law for both intensity and polarization. The first case was employed previously by several authors as it has an analytical solution. Here we compare this solution with the Rayleigh scattering calculation and demonstrate that both cases result in very similar shapes of the polarization curves.

By considering the finite size of the host star we have to take into account that different parts on the stellar surface have in general individual distances to the planet because of the spherical shape of the star. In our calculations the surface of the star is divided into individual pixels. They are chosen to be small enough so that all light rays emitted from one stellar surface element arrive nearly parallel on the planetary atmosphere, which requires our assumption that the planet is small compared to the star. Therefore the incident radiation from one single stellar pixel results in a contribution to the scattered radiation that can be computed as in the case of a point-like star. In general, the incident flux, the scattering angle, and the phase angle differ for every pixel. The total scattered radiation field is obtained simply by adding up the contributions from all stellar surface elements. It is crucial to consider only the actually visible stellar surface as seen from the position of the planet, which approaches exactly one stellar hemisphere for very distant planets, but is in general smaller than that for close-by planets.

The ideal size of the stellar pixel in principle depends on the stellar radius and the size of the planetary orbit. For the calculations presented in this paper we chose a pixel size of one degree both in longitude and latitude. Pixel sizes of more than five degrees have occasionally lead to inaccurate results in test calculations. For smaller pixels the computed flux and polarization were always independent of the pixel size, as long as it is still high enough to prevent numerical errors due to the machine precision.
At the end of this section we briefly review the results for a point-like host star compared with an extended star for completeness' sake.


\subsection{Lambert-Rayleigh approximation}
\label{subsec:lambert}

The flux $F$ from the unresolved planetary system measured by a distant observer consists of two parts, namely
\begin{equation}\label{eq:totFlux}
F = F_\star + F_\mathrm{sc}
\rm \ ,
\end{equation}
where $F_\star$ is the direct stellar flux and $F_\mathrm{sc}$ represents the fraction of the stellar flux which is scattered by the extrasolar planet into the line-of-sight. We neglected the thermal emission from the planet because we are mainly interested in the optical wavelengths at this stage.

The direct flux is given by
\begin{equation}\label{eq:directFlux}
F_\star = 
2\pi \frac{R_\star^2}{d^2} \int_0^1{I\mu d\mu}
\rm \ ,
\end{equation}
with $R_\star$ the stellar radius, $d$ the distance from the star to the observer that is assumed to be very large compared to the semi-major axis of the planetary orbit, $I$ the intensity at the stellar surface, and $\mu\!=\!\cos\theta$, $\theta$ being the angle between the line-of-sight and a stellar surface element as seen from the center of the star. The integral accounts for limb darkening too. The absolute value of the flux is of no concern for the polarization, which is a relative quantity. We define it by choosing the normalization condition
\begin{equation}\label{eq:normalization}
2\pi R_\star^2 \int_0^1{I\mu d\mu} = 1
\rm \ ,
\end{equation}
which scales the intensity at the surface of the star. The direct stellar flux then simplifies to $F_\star\!=\!1/d^2$. We have modeled the limb darkening with the expression introduced by \citet{claret2000}
\begin{equation}\label{eq:limbdarklaw}
\frac{I(\mu)}{I(1)} = 1 - \sum_{k=1}^4 a_k \left(1-\mu^\frac{k}{2}\right)
\rm \ .
\end{equation}
In this paper we used two sets of coefficients, one for a hypothetical exoplanet orbiting a giant star ($a_1\!=\!0.4908$, $a_2\!=\!-0.5346$, $a_3\!=\!1.2703$, $a_4\!=\!-0.2889$) and another one for the Sun ($a_1\!=\!0.4767$, $a_2\!=\!-0.1591$, $a_3\!=\!1.0711$, $a_4\!=\!-0.5154$), both approximating the limb darkening in the B passband \citep{claret2000}.

To obtain the scattered flux $F_\mathrm{sc}$ we considered the total energy incident on the planet per second and frequency unit from one single stellar surface element
\begin{equation}\label{eq:incidentFlux}
F_\mathrm{inc,p} = \pi r^2 I_\mathrm{p} \, \mathrm{d}{\mathit{\Omega}}_\mathrm{p}
\rm \ ,
\end{equation}
where the index ``p'' indicates that the corresponding quantity depends on the selected stellar surface element or pixel. The parameter $r$ represents the radius of the planet, i.e.\ $\pi r^2$ is just the cross-section of the planet, $I_\mathrm{p}$ is the intensity emitted from the surface element in the direction of the planet, and d${\mathit{\Omega}}_\mathrm{p}$ is the solid angle covered by the stellar surface element as seen from the planet and should not be confused with the position angle of the ascending node $\Omega$. The distance from the stellar surface element to the planet is implicitly contained within the solid angle d${\mathit{\Omega}}_\mathrm{p}$ and thus differs for different pixels e.g.\ at the limb or disk center. If the planet is small compared to the star, we can assume that d${\mathit{\Omega}}_\mathrm{p}$ coincides for all positions on the planet. Correspondingly, we compute d${\mathit{\Omega}}_\mathrm{p}$ always for the center of the mass of the planet. From the incident light $F_\mathrm{inc,p}$ a fraction
\begin{equation}\label{eq:fractionScattered}
f_\mathrm{sc,p} = \frac{1}{\pi} p \, \Phi(\alpha_\mathrm{p})
\end{equation}
is scattered towards the observer. Here $1/\pi$ represents the fraction of light scattered back to the star by a Lambert disk at phase angle zero, $p$ is the geometric albedo, which is 2/3 for a Lambert sphere, and $\alpha_\mathrm{p}$ is the phase angle, i.e.\ the angle between a stellar surface element and the observer as seen from the planet. For a Lambert sphere the phase function $\Phi$ takes the analytical form \citep{russell1916}
\begin{equation}\label{eq:phaseFunction}
\Phi(\alpha_\mathrm{p}) = 
\frac{1}{\pi}\left( \sin\alpha_\mathrm{p} + ( \pi - \alpha_\mathrm{p} )
                    \cos\alpha_\mathrm{p} \right)
\rm \ .
\end{equation}
Combining Eqs.~(\ref{eq:incidentFlux}) and (\ref{eq:fractionScattered}) we find the contribution to the flux that is scattered by the planet towards a distant observer
\begin{equation}\label{eq:scatFlux}
F_\mathrm{sc} = \frac{pr^2}{d^2} 
                  \int_\mathrm{visible} \Phi(\alpha_\mathrm{p})
                                        I_\mathrm{p} \, \mathrm{d}{\mathit{\Omega}}_\mathrm{p}
\rm \ .
\end{equation}
The integral is carried out over the visible surface of the star as seen from the planet. It is easily verified that the scattered flux $F_\mathrm{sc}$ as given in Eq.~(\ref{eq:scatFlux}) approaches the simplified case of the star as a point-source for large planetary orbits, i.e.\ the second part given in Eq.~(\ref{eq:scatFluxPoint}).

Scattering polarization was added in an ad hoc fashion, because a Lambert sphere scatters the incident radiation isotropically and in principle unpolarized. We estimated the polarization in a similar way as was done by \citet{seageretal2000} for a Lambert sphere, thus creating an idealized case, which is sufficient to discuss the influence of orbital parameters on the polarization curves. The calculated degree of polarization defines an upper limit, so that an inferred planetary radius corresponds to a lower limit.

We employed two different reference frames to represent the Stokes $Q$ and $U$ parameters. The final results are all given with respect to celestial north for the observer on Earth, so that a positive Stokes $Q$ describes light that is linearly polarized parallel to the local meridian of the observer. This first reference frame is referred to as the ``observer's frame''. A different reference frame is temporarily used while adding up contributions from different stellar surface elements, where we define the Stokes parameters with respect to the scattering plane so that positive Stokes $Q$ represents light that is linearly polarized perpendicular to the scattering plane.

Here we assumed that light incident on the planet is scattered and that the resulting Stokes $Q$ and $U$ are defined by Rayleigh scattering. Considering scattered light originating from one stellar surface element, and choosing the scattering plane as the reference frame (indicated by primed variables), we obtained a contribution to Stokes $Q'$ given by 
\begin{equation}\label{eq:Qpixel}
Q'_\mathrm{p} = \frac{1}{d^2} F_\mathrm{inc,p} \, f_\mathrm{sc,p} 
                \sin^2\psi_\mathrm{p}
              = \frac{pr^2}{d^2} \, \Phi(\alpha_\mathrm{p})
                I_\mathrm{p} \, \mathrm{d}{\mathit{\Omega}}_\mathrm{p} \sin^2\psi_\mathrm{p}
\rm ,
\end{equation}
while $U'_\mathrm{p}\!=\!0$. The scattering angle $\psi_\mathrm{p}$ is related to the phase angle by $\alpha_\mathrm{p}=\pi-\psi_\mathrm{p}$. The total Stokes $Q$ and $U$ parameters measured by the distant observer in the observer's frame is obtained by transforming the reference frame and integrating over the visible surface of the star (as seen from the planet):
\begin{equation}\label{eq:totalQU}
\left[ 
\begin{array}{c}
	Q \\ U
\end{array}
\right]
=
\int_\mathrm{visible} \vec{L}(\beta_\mathrm{p})
\left[
\begin{array}{c}
	Q'_\mathrm{p} \\ 0
\end{array}
\right]
\rm \ .
\end{equation}
The Mueller rotation matrix $\vec{L}$ is defined by \citep[e.g.][]{stenflo1994}
\begin{equation}\label{eq:rotMatrix}
\vec{L}(\beta_\mathrm{p}) = 
\left[
\begin{array}{cc}
	\cos2\beta_\mathrm{p}  & \sin2\beta_\mathrm{p} \\
	-\sin2\beta_\mathrm{p} & \cos2\beta_\mathrm{p}
\end{array}
\right]
\rm \ ,
\end{equation}
where $\beta_\mathrm{p}$ is the rotation angle to transform the scattering plane reference system into the observer's frame. It depends on the stellar surface element. Note that Stokes $V$ is zero in both reference frames because we assumed that the light emitted from the star is unpolarized. The polarization induced by the unresolved extrasolar planet is then determined by $Q/F$ and $U/F$. We note that non-Rayleigh types of scattering influence the shape of the polarization curves \citep{seageretal2000}. We will consider these cases in a forthcoming paper.


\subsection{Rayleigh scattering}\label{subsec:rayleigh}

Let us now consider the case when both the intensity and polarization of the light scattered in the planetary atmosphere are distributed according to the Rayleigh law. This implies solving a self-consistent radiative transfer problem. Here we solved this problem under the assumptions (in addition to those mentioned in Sect.~\ref{subsec:assumptions}) that
\begin{itemize}
	\item the atmosphere is plane-parallel and static and consists of homogeneous layers,
	\item the planet is spherically symmetric,
	\item the radiation entering the atmosphere is unidirectional and unpolarized,
	\item an incoming photon is either absorbed or scattered in the continuum,
	\item an absorbed photon does not alter the atmosphere (model atmosphere includes effects of irradiation),
	\item single scattering approximation (a photon might still be extinct after scattering).
\end{itemize}
Thus we solve a standard radiative transfer equation for scattered polarized radiation of a given frequency
\begin{equation}\label{eq:rt}
\mu\frac{d\vec{I}}{d\tau} = \vec{I} - \vec{S}
\end{equation}
with the total source function
\begin{equation}\label{eq:sourcef}
\vec{S} = \frac{\kappa_\mathrm{c}\vec{B}+\sigma_\mathrm{R}\vec{S}_\mathrm{R}}{\kappa_\mathrm{c}+\sigma_\mathrm{R}}
\rm \ ,
\end{equation}
where $\kappa_\mathrm{c}$ and $\sigma_\mathrm{R}$ are opacities due to continuum absorption and Rayleigh scattering, while $\vec{S}_\mathrm{R}$ and $\vec{B}$  are the Rayleigh source function and the unpolarized thermal emission, respectively.

The Rayleigh source function $\vec{S}_\mathrm{R}$ is expressed as usual via the Rayleigh phase matrix $\vec{\hat{P}}(\mu,\mu')$, depending on the directions of the incident and scattered light. It has contributions from scattering of both the intrinsic thermal emission and incident stellar light. However, in the blue part of the spectrum, where the polarization signal due to Rayleigh scattering is the largest, the intensity of the thermal emission is negligible compared to that of the reflected light. This is in contrast to the assumption by \citet{sengupta2008} who neglected the polarization of the reflected light and took only into account the polarization of the thermal emission to interpret measurements in the B passband. In this case he had to also assume a non-negligible oblateness of the planet. This approach can only be justified to explain polarization in red/infrared wavelengths which has not yet been detected. 

Taking into account that the incident stellar radiation is unidirectional (in case of a finite-size star the radiation comes from a given stellar pixel), we obtain
\begin{equation}\label{eq:sourcef_Ray}
\vec{S}_\mathrm{R}(\tau,\mu') = \frac{1}{4\pi}\vec{\hat{P}}(\mu,\mu')\vec{I}(\tau,\mu)
\rm \ ,
\end{equation}
where for single scattering
\begin{equation}\label{eq:singlescat}
\vec{I}(\tau,\mu)=\vec{I}_\star e^{-\frac{\tau}{\mu}}
\rm \ ,
\end{equation}
and $\tau$ is the optical depth of the layer in the planetary atmosphere from $\tau=0$ at the top. In the same way the scattered radiation is reduced due to the optical depth on the way out of the atmosphere.

The Stokes vector of the reflected light is obtained by solving the radiative transfer problem as described above for a given vertical distribution of the temperature and opacity in a planetary atmosphere. The radiation flux is then obtained by integrating the Stokes vector over the illuminated planetary surface with a coordinate grid on the planetary surface. Our tests have shown that a grid of 6\deg$\times$6\deg\ is sufficient to achieve the necessary accuracy. Thus the phase function $\vec{\Phi}(\alpha)$ is in this case numerically evaluated.

Our model includes the following opacity sources: (i) Rayleigh scattering on H, H$_2$, He, H$_2$O, CH$_4$, and Thomson scattering on electrons, with all scattering species contributing to polarization, and (ii) continuum absorption due to H, H$^-$, H$_2$$^+$, H$_2$$^-$, He, He$^-$, Si, Mg, and Fe. The number densities of the species were calculated with a chemical equilibrium code described in \cite{berdyuginaetal2003}. For an example calculation presented in Sect.~\ref{subsec:comparison} we employed standard model atmospheres by \citet{allardetal2001}. For interpreting real data, models of irradiated atmospheres are to be used.


\subsection{A comparison between the two cases}\label{subsec:comparison}

The difference between the above two cases is in the angular distribution of the intensity of the scattered light: in the first case, it is assumed to be according to the Lambert sphere law, while in the second case, it is according to the Rayleigh law. Consequently the corresponding phase functions $\vec{\Phi}_\mathrm{L}(\alpha)$ and $\vec{\Phi}_\mathrm{R}(\alpha)$  are certainly different, and it is important to evaluate the resulting difference in polarization curves.

\begin{figure}
  \resizebox{\hsize}{!}{\includegraphics{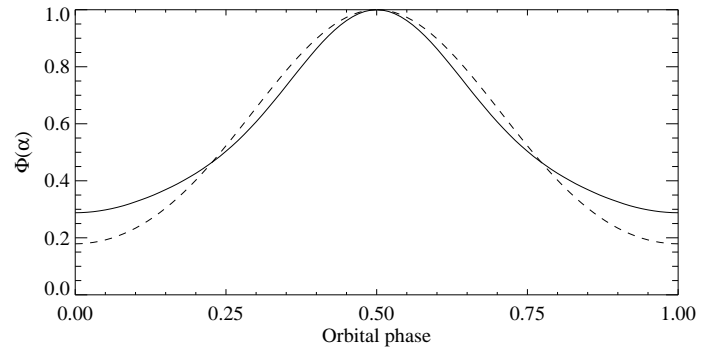}}
  \caption{Normalized phase functions for the Lambert sphere (dashed line) and  the Rayleigh law (solid line). The calculation is for a circular planetary orbit with the inclination $i\!=\!30$\deg, and in case of Rayleigh scattering for the wavelength of 4500~\AA. The differences occurring in the gradients of the two functions are caused by different angular dependences of the Rayleigh scattering law and a Lambert sphere. For higher inclinations, the difference in gradients is slightly modified. The dependence on the wavelength is a second order effect. The two curves have been renormalized to one at the maximum to make the comparison of the shapes more apparent.}
  \label{fig:LR_phasefunc}
\end{figure}

\begin{figure}
  \resizebox{\hsize}{!}{\includegraphics{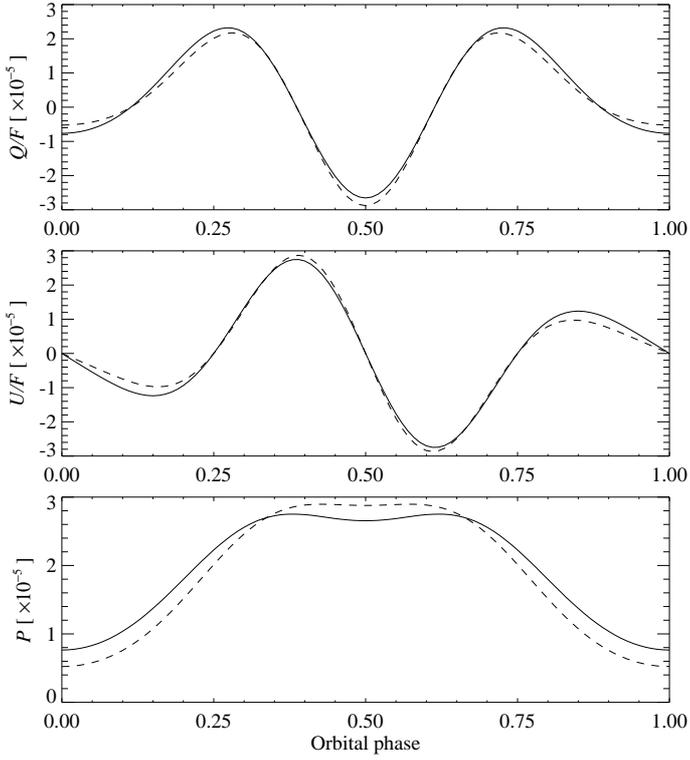}}
  \caption{Polarization curves obtained with the phase functions presented in Fig.~\ref{fig:LR_phasefunc}. The scale of the Rayleigh law curves has been adjusted to fit the Lambert sphere curves. The differences between the shapes of Stokes $Q$ and $U$ are negligible given the current best accuracy of polarimetric measurements. Thus the Lambert sphere approximation for the intensity of the reflected light is justified for this case.}
  \label{fig:LR_polcurves}
\end{figure}

The shapes of the functions $\vec{\Phi}_\mathrm{L}(\alpha)$ and $\vec{\Phi}_\mathrm{R}(\alpha)$ (normalized to their maxima) are shown in Fig.~\ref{fig:LR_phasefunc}. Their behavior with the orbital phase (which differs from the phase angle $\alpha$) is as expected: the maximum illuminated area is observed at the exterior conjunction (``full moon", phase 0.5), while the minimum is at the interior one (``new moon", phase 0.0). Some differences between the two functions are seen in their gradients, which are due to the different angular dependencies of Rayleigh scattering and a Lambert sphere. The Rayleigh phase function also depends on the wavelength, but this effect is significantly smaller. Note that we normalized the plotted phase functions to one at their maxima to make the comparison of the shapes more apparent. In reality the two curves differ somewhat in scale depending on the detailed properties of the model atmosphere employed in the Rayleigh scattering case, with negligible influence on polarization.

The polarization curves resulting from these phase functions are shown in Fig.~\ref{fig:LR_polcurves}. The overall scale of polarization obviously differs for the two cases. Therefore we have changed the scale of the Rayleigh law results for an easy comparison of the curve shapes. The shapes of the polarization curves do not differ by a measurable amount for the case considered in this paper, i.e., when the reflected light is significantly dimmer than the direct stellar light and the planet remains spatially unresolved. This is because of the normalization of the Stokes parameters to the total flux from the system. For resolved planets, the differences between the shapes of the polarization curves of the Rayleigh and Lambert cases increase but still remain small. This renders the Lambert sphere law a good approximation for the intensity of the reflected light, especially when obtaining orbital parameters of planets, which is the primary goal of our case study. To determine the composition and thermodynamic properties of the planetary atmosphere, a self-consistent polarized radiative transfer calculation as presented in Sect.~\ref{subsec:rayleigh} is required.


\subsection{Host star as a point-source}\label{subsec:point}

We will compare the model described above for an extended host star with the simplified case, in which the star is treated as a point source. This case is applicable if the semi-major axis of the planetary orbit exceeds the radius of the star, so that all light rays from the star can be assumed to arrive parallel on the planet. Hence the total flux at the observer with the normalization condition in Eq.~(\ref{eq:normalization}) is given by
\begin{equation}\label{eq:scatFluxPoint}
F = \frac{1}{d^2} + \frac{pr^2}{d^2D^2} \, \Phi(\alpha) 
\rm \ ,
\end{equation}
where $D$ is the distance between the star and the planet. The two parts in Eq.~(\ref{eq:scatFluxPoint}) correspond to the direct stellar flux and the flux scattered by the planet, respectively. Since the incident radiation arrives parallel on the planet, it is sufficient to consider only one phase angle $\alpha$. By comparing Eq.~(\ref{eq:scatFluxPoint}) with the extended star model we find that $D$ does not appear explicitly in Eq.~(\ref{eq:scatFlux}) because it is implicitly contained there within the solid angle d${\mathit{\Omega}}_\mathrm{p}$ and in general differs for every individual surface element. In the limit of very large planetary orbits, i.e.\ approaching the point-source model, the distances between the planet and different stellar surface elements approach the same value, namely $D$, and can be explicitly expressed in Eq.~(\ref{eq:scatFluxPoint}).

With respect to the scattering plane Stokes $U'\!=\!0$ and
\begin{equation}\label{eq:QUpoint}
Q' = \frac{pr^2}{d^2D^2} \, \Phi(\alpha) \sin^2\psi
\rm \ ,
\end{equation}
which can be converted to the observer's frame with the Mueller rotation matrix $\vec{L}$ given in Eq.~(\ref{eq:rotMatrix}).


\section{Parameter study of polarization phase curves}\label{sec:lightcurves}

We perform in this section a parameter study and discuss the influence of different parameters on the $Q/F$ and $U/F$ phase curves and on the degree of polarization $P\!=\!\sqrt{Q^2+U^2}/F$. Since we showed in Sect.~\ref{subsec:comparison} that polarization curves calculated with the Lambert-Rayleigh approximation do not differ from those based on the self-consistent radiative transfer case for unresolved systems, here we employ the first case for simplicity.

We considered the inclination of the planetary orbit $i$, the position angle of the ascending node $\Omega$, the eccentricity $e$, and the longitude of the periastron $\omega$. In addition we compared models with extended and point-like stars to determine the parameter range for which the latter model serves as a good approximation.

Most features of the polarized phase curves can be understood in terms of only a few quantities (cf.\ Eq.~(\ref{eq:QUpoint})):
\begin{enumerate}
\item The polarization depends on the scattering angle $\psi$: it is the largest for 90\deg\ scattering at elongation and drops to zero for backward and forward scattering due to geometrical reasons.
\item The greater the phase function $\Phi$, the larger the polarization, since $\Phi$ describes the fraction of the incident light scattered towards the observer. The phase function reaches the maximum at the smallest phase angle when the planet is farthest away from the observer and the minimum at the largest phase angle when the planet is closest to the observer.
\item The polarization scales with the inverse square of the distance $D$ from the star to the planet, in accordance with the incident flux. In the case of elliptical orbits $D$ varies with the orbital phase.
\item The larger the radius of the planet, the larger the scattering surface and polarization.
\item The linear polarization is always directed perpendicular to the scattering plane, i.e.\ perpendicular to the direction from the star to the planet as seen in projection on the sky (we neglect here effects of magnetic fields). Therefore, the linear polarization generally rotates while the planet orbits the star, so that Stokes $Q$ and $U$ are exchanged and change their signs.
\end{enumerate}
Both the radius of the planet and the semi-major axis of the planetary orbit only scale the amplitude of the polarized phase curves, and we will not further discuss them in this section. Uncertainties in the planetary radius do not influence the orbital solution. We assume $a\!=\!0.05$~AU, $r\!=\!1$~$R_\mathrm{J}$ (Jupiter radius), and $R_\star\!=\!1$~$R_\mathrm{\odot}$. All other points in the list, including also the orbital parameters $i$, $\Omega$, $e$, $\omega$, modify the shape of the light curves. 


\subsection{Inclination of the planetary orbit}\label{subsec:inclination}

First we discuss the inclination $i$ of the planetary orbit. It is defined in a way that the values between 0\deg\ and 90\deg\ correspond to the counter-clockwise rotation of the planet as seen in projection by the observer, while angles between 90\deg\ and 180\deg\ indicate the clockwise rotation. The orbit is seen edge-on at an inclination $i\!=\!90$\deg.

\begin{figure}
  \resizebox{\hsize}{!}{\includegraphics{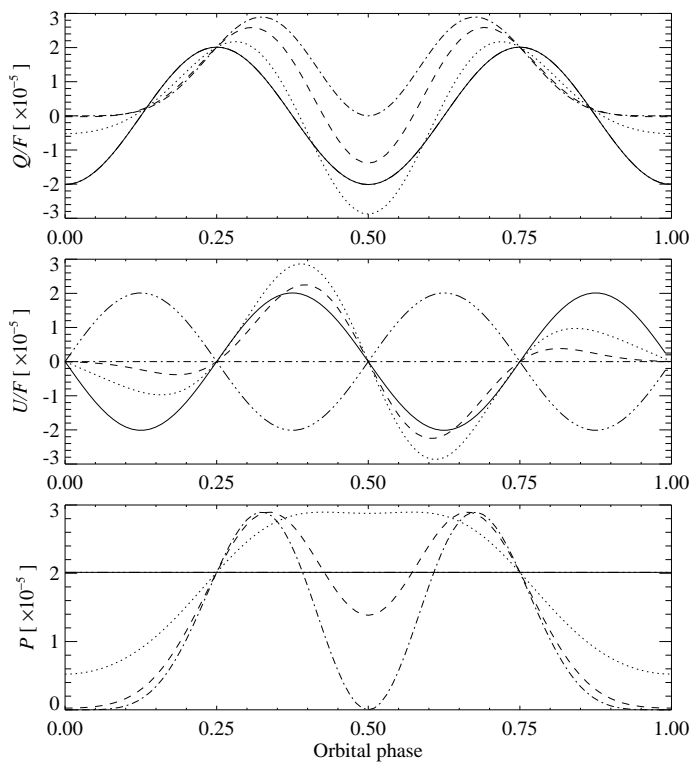}}
  \caption{Polarization curve dependence on inclination of the planetary orbit. The inclinations cover $i\!=\!0$\deg\ (solid), $i\!=\!30$\deg\ (dotted), $i\!=\!60$\deg\ (dashed), $i\!=\!90$\deg\ (dash-dotted), and $i\!=\!180$\deg\ (dash-triple-dotted). The curves for $i\!=\!0$\deg\ and $i\!=\!180$\deg\ coincide in the top and bottom panels and are shown with solid lines. Further parameters: $a\!=\!0.05$~AU, $\Omega\!=\!90$\deg, $e\!=\!0$, $r\!=\!1$~$R_\mathrm{J}$, and $R_\star\!=\!1$~$R_\mathrm{\odot}$. The orbital phase zero is defined by the largest phase angle (and by a position angle of 180\deg\ in the special cases of $i\!=\!0$\deg\ and $i\!=\!180$\deg).}
  \label{fig:inclination}
\end{figure}

Figure~\ref{fig:inclination} illustrates how the inclination $i$ affects the linear polarization at different times of the planet period, for simplicity in the case of a circular orbit. Basically it dictates how strongly the extrema of the polarization curves are pronounced. As always for circular orbits, we defined the time zero $t_0$ by the passage of the planet through the largest phase angle, i.e\ when it is closest to the observer. The orbital phase represents the fractional time passed within one orbit relative to $t_0$.

When looking face-on onto the orbit, i.e.\ with $i\!=\!0$\deg\ or $i\!=\!180$\deg, the degree of polarization remains constant because the planet is always seen under the same phase angle. Still the rotation of the linear polarization leads to a sine-like variation of Stokes $Q$ and $U$ with the same amplitudes, but a phase shift of 90\deg.

The closer to edge-on we observe the orbit, the closer Stokes $Q$, Stokes $U$, and polarization approach zero for the orbital phase 0.0. This minimum in polarization is very broad because the phase function and $\sin^2\psi$ (forward scattering) are simultaneously small. Interestingly, both $Q/F$ and $U/F$ are usually very small near $t_0$, except for inclinations below 30\deg, a helpful feature for the determination of the zero point of the polarization scales (cf.\ Sect.~\ref{sec:fits}). The polarization maxima occur near maximum elongations, where $\sin^2\psi\!=\!1$, but slightly shifted towards smaller phase angles (i.e.\ closer to orbital phase $\varphi\!=\!0.5$), because for those phases the illuminated fraction of the planetary disk increases for the observer towards $\varphi\!=\!0.5$.

For the specific choice of $\Omega\!=\!90$\deg\ in Fig.~\ref{fig:inclination} all $Q/F$ curves are symmetric with respect to phase $\varphi\!=\!0.5$, while all $U/F$ curves are antisymmetric. This indicates that the shape of $Q/F$ is independent of the direction into which the planet moves within the orbit and thus identical for orbits with inclinations $i$ and 90\deg$-i$. On the contrary, $U/F$ changes its sign for opposite rotation directions (cf.\ the solid and dash-triple-dotted curves in Fig.~\ref{fig:inclination}). Note however note that these features depend on $\Omega$.

We also found that the extrema shift with inclination, not so much in the total degree of polarization, but clearly in $Q/F$ and $U/F$. If we look for example at the $i\!=\!60$\deg\ case (dashed curve, Fig.~\ref{fig:inclination}), we find that the two main extrema of $U/F$ occur relatively close to phase $\varphi\!=\!0.5$, even closer than the two maxima of $Q/F$, which is obvious from geometrical considerations. As we will clarify in the subsection below these properties of $Q/F$ and $U/F$ can be exchanged by variations of the position angle of the ascending node $\Omega$.

Our model will allow us to significantly improve the mass determination of extrasolar planets, for most of which only a lower limit $M\sin i$ is known. According to the standard model of planet formation we expect that the angle between the inclination of the planetary orbit and the spin axis of the host star is small as in the case of the solar system. Indeed, it was found to be less than 11\deg\ in the sample of three transiting exoplanets for which the alignment was determined \citep{winnetal2005,winnetal2006,wolfetal2007}. Therefore the measurement of the polarization curves, which are very sensitive to the inclination of the orbit as discussed above, represents a powerful method to constrain also the inclination $i_\star$ of the stellar spin axis to within a few degrees.


\subsection{Position angle of the ascending node}\label{subsec:OMEGA}

\begin{figure}
  \resizebox{\hsize}{!}{\includegraphics{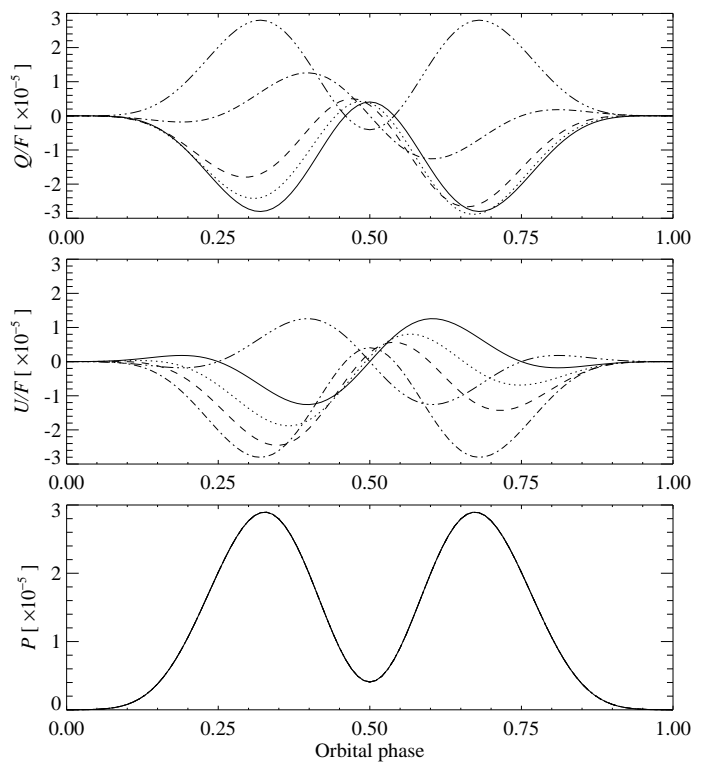}}
  \caption{Polarization curve dependence on the position angle of the ascending node $\Omega$. Different curves correspond to $\Omega\!=\!0$\deg\ (solid), $\Omega\!=\!10$\deg\ (dotted), $\Omega\!=\!20$\deg\ (dashed), $\Omega\!=\!45$\deg\ (dash-dotted), and $\Omega\!=\!90$\deg\ (dash-triple-dotted). Further parameters: semi-major axis $a\!=\!0.05$~AU, inclination $i\!=\!75$\deg, eccentricity $e\!=\!0$, radius of the planet $r\!=\!1$~$R_\mathrm{J}$, and a stellar radius $R_\star\!=\!1$~$R_\mathrm{\odot}$. The orbital phase zero is defined by the largest phase angle.}
  \label{fig:OMEGA}
\end{figure}

A variation of the position angle of the ascending node $\Omega$ rotates the orbit around the line-of-sight and is analogous to the rotation of the reference frame for polarization. The degree of polarization remains unaffected accordingly, but the plane of linear polarization rotates, as shown in Fig.~\ref{fig:OMEGA}. While increasing $\Omega$ from 0\deg\ (solid) to 45\deg\ (dash-dotted) the shape of $Q/F$ changes from being symmetric to antisymmetric (with respect to $\varphi\!=\!0.5$) and effectively takes over the properties of the $U/F$ curve. The opposite is true for $U/F$. At $\Omega\!=\!90$\deg\ (dash-triple-dotted) both curves again reach the same shape as for $\Omega\!=\!0$\deg, but with opposite sign.

The polarization curves contain one ambiguity, since they remain unaltered when shifting $\Omega$ by 180\deg. This is of course an inherent property of polarization, because Stokes $Q$ and $U$ are not subject to change under a rotation of the plane of polarization by 180\deg.


\subsection{Elliptic orbits}\label{subsec:eccentricity}

\begin{figure}
  \resizebox{\hsize}{!}{\includegraphics{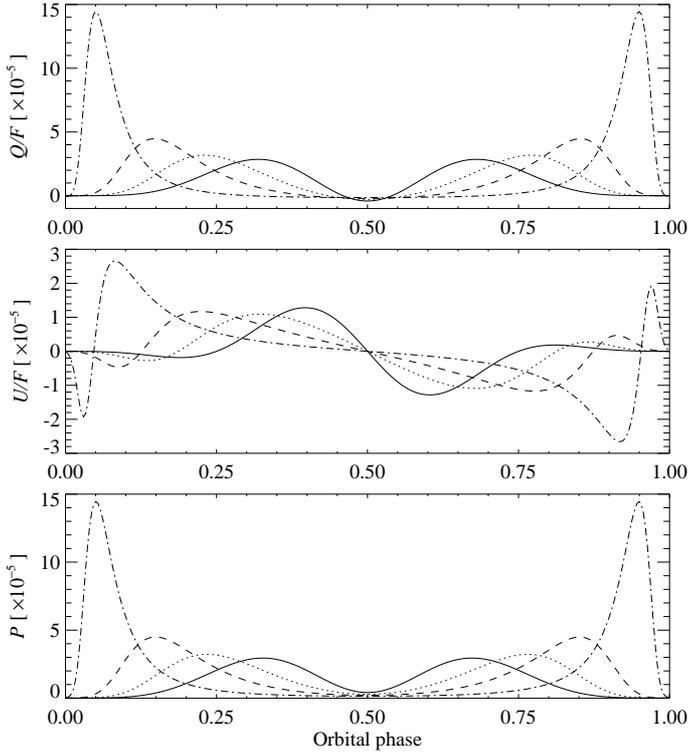}}
  \caption{Polarization curve dependence on the eccentricity. Different curves correspond to $e\!=\!0.0$ (solid), $e\!=\!0.2$ (dotted), $e\!=\!0.4$ (dashed), and $e\!=\!0.7$ (dash-dotted). Further parameters: $a\!=\!0.05$~AU, $i\!=\!75$\deg, $\Omega\!=\!90$\deg, $\omega\!=\!90$\deg, $r\!=\!1$~$R_\mathrm{J}$, and $R_\star\!=\!1$~$R_\mathrm{\odot}$. The orbital phase zero is defined by the largest phase angle, i.e.\ when the planet is closest to the observer, which in this particular case coincides with the periastron epoch due to our particular choice of $\omega\!=\!90$\deg.}
  \label{fig:ecc}
\end{figure}

\begin{figure}
  \resizebox{\hsize}{!}{\includegraphics{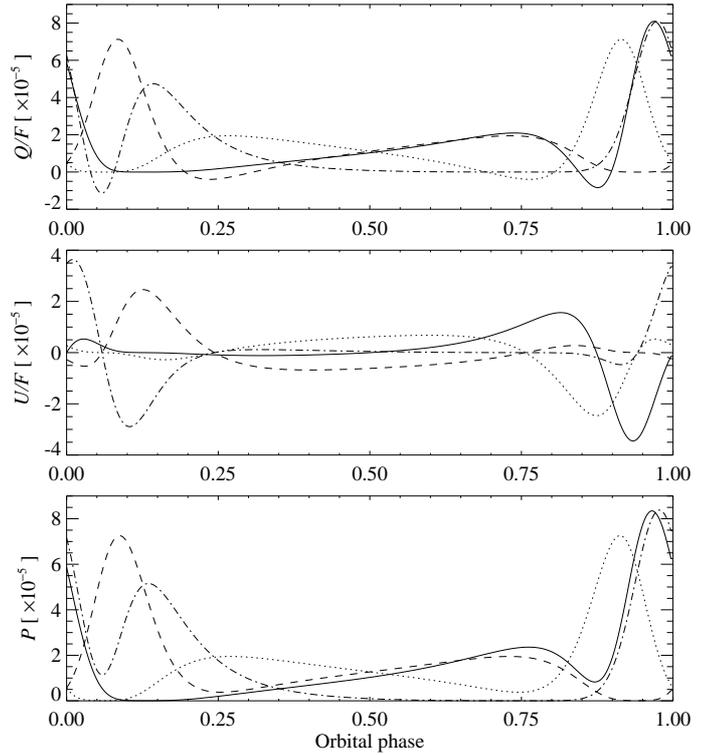}}
  \caption{Polarization curve dependence on the periastron longitude for eccentric orbits. Different curves correspond to $\omega\!=\!0$\deg\ (solid), $\omega\!=\!45$\deg\ (dotted), $\omega\!=\!135$\deg\ (dashed), and $\omega\!=\!225$\deg\ (dash-dotted). Further parameters: $a\!=\!0.05$~AU, $i\!=\!75$\deg, $\Omega\!=\!90$\deg, $e\!=\!0.4$, $r\!=\!1$~$R_\mathrm{J}$, and $R_\star\!=\!1$~$R_\mathrm{\odot}$. The orbital phase zero is defined by the periastron passage.}
  \label{fig:periastron}
\end{figure}

In the case of eccentric orbits we need to distinguish two additional parameters that affect the polarization curves in a different manner, namely the eccentricity $e$ and the periastron longitude $\omega$.

For eccentric orbits we choose the periastron passage as zero $t_0$, because the polarization peaks shift towards the periastron, and the radial velocity method well constrains the periastron epoch for elliptic orbits. To allow a comparison with circular orbits we selected $\omega\!=\!90$\deg\ in Fig.~\ref{fig:ecc} so that the periastron epoch coincides with $t_0$ used in Sects.~\ref{subsec:inclination} and \ref{subsec:OMEGA}, where it was defined by the largest phase angle.

The eccentricity $e$ influences the polarization curves in two ways when the semi-major axis is kept fixed: it shifts and scales the extrema of the light curves (Fig.~\ref{fig:ecc}). The shift of the extrema results from the varying orbital velocity of the planet. With increasing eccentricity the extrema group closer to the periastron passage, although maximum polarization is generally not reached exactly at the periastron. At the same time most extrema become stronger, in the case shown in Fig.~\ref{fig:ecc} even all extrema, because the planet approaches the star much closer at periastron for eccentric orbits (with fixed semi-major axis). Recall that the incident flux on the planet and thus the scattering polarization scale with the inverse square of the distance $D$ between the star and the planet.

Initially it might sound surprising that all extrema strengthen for elliptic orbits in Fig.~\ref{fig:ecc}, since for circular orbits the maximum polarization always occurs at phase angles smaller than 90\deg, which occur when the planet moves through the part of the orbit around the apastron for our specific choice of $\omega\!=\!90$\deg. However, for very eccentric orbits the phase angle decreases quickly after the periastron passage, and the phase function exceeds the value of 0.5 already when the planet moves through the descending node. This happens when $D$ is still much smaller than the semi-major axis and long before the maximum elongation. Moreover the phase angles at which the extrema occur move closer to the periastron due to the strong dependence of polarization on the distance $D$. The distance dependence of the polarization also explains why the extrema in Fig.~\ref{fig:ecc} are strongly scaled up near the periastron passage. By comparing the circular orbit (solid) with the most eccentric case (dash-dotted) we find indeed that the inconspicuous minimum of $U/F$ at $\varphi\!=\!0.2$ in the circular case is shifted to $\varphi\!=\!0.03$ and scaled up by about a factor of 10, the maximum in $Q/F$ at $\varphi\!=\!0.3$ (circular case) is increased by a factor of 5, and the maximum of $U/F$ at $\varphi\!=\!0.4$ (circular case) is scaled up only by about a factor 2.

In Fig.~\ref{fig:periastron} we demonstrate the effect of the longitude of the periastron $\omega$, which in principle rotates and positions the elliptic orbit within the orbital plane. This results in an enhancement of the polarization curve extrema that lie closest to the periastron epoch. Therefore some Stokes $U$ curves in Fig.~\ref{fig:periastron} exhibit pronounced negative peaks that are not apparent for other values of $\omega$. On the other hand, a variation of $\omega$ can not produce negative (or positive) polarization if it is absent in the corresponding circular case with the same orbital orientation. Such an example is visible in Stokes $Q$ of Fig.~\ref{fig:periastron}, where strong peaks with negative $Q$ are nonexistent for any $\omega$, as expected from the corresponding circular orbit (solid line, top panel, Fig.~\ref{fig:ecc}).

From Fig.~\ref{fig:periastron} we find some symmetry properties that occur with a change of $\omega$. The curves with $\omega\!=\!90$\deg$\,\pm\gamma$ (or $\omega\!=\!270$\deg$\,\pm\gamma$), with $\gamma$ any angle, are just mirrored at $\varphi\!=\!0.5$ in the total degree of polarization and in Stokes $Q$, while in Stokes $U$ they are mirrored at $\varphi\!=\!0.5$ combined with a sign change (see dotted and dashed curves). In Stokes $Q$ and $U$ this symmetry relation depends on the particular choice of $\Omega$, but it holds for any $\Omega$ in the degree of polarization.


\subsection{Size of the host star and secondary eclipses}\label{subsec:extended}

\begin{figure}
  \resizebox{\hsize}{!}{\includegraphics{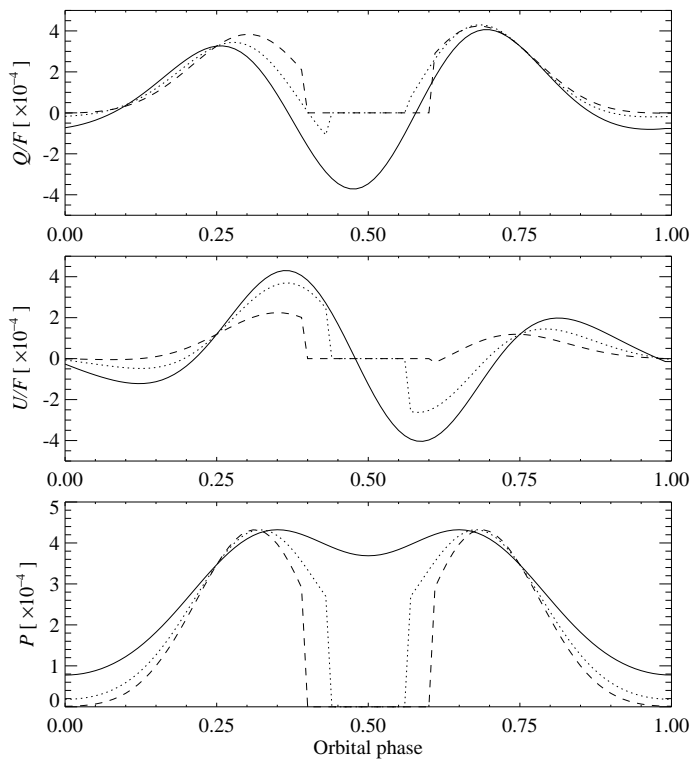}}
  \caption{Influence of secondary eclipses. The curves correspond to three different inclinations: $i\!=\!40$\deg\ (solid), which is too small for the occurrence of eclipses, $i\!=\!60$\deg\ (dotted), and $i\!=\!80$\deg\ (dashed). The additional parameters correspond to a hypothetical hot Jupiter orbiting a giant with $a\!=\!0.109$~AU, $\Omega\!=\!100$\deg, $e\!=\!0$, $r\!=\!9$~$R_\mathrm{J}$, and a stellar radius $R_\star\!=\!13.3$~$R_\mathrm{\odot}$. The orbital phase zero is defined by the largest phase angle. For other stellar or planetary radii the curves are scaled in the degree of polarization, but their shapes remain the same.}
  \label{fig:eclipses}
\end{figure}

Previous simulations of the polarization signal from exoplanets always assumed that the light from the parent star falls parallel onto the planet (even in the case of hot Jupiters), a model we will refer to as ``point-like'' in the following. However, the assumption of parallel rays incident on the planet clearly fails in the case of hot Jupiters orbiting giants. 
This necessitates a correct modeling with an extended, finite-sized host star, which takes into account the strong distance dependence of the resulting polarization.

\begin{figure*}
  \centering
  \includegraphics[width=17cm]{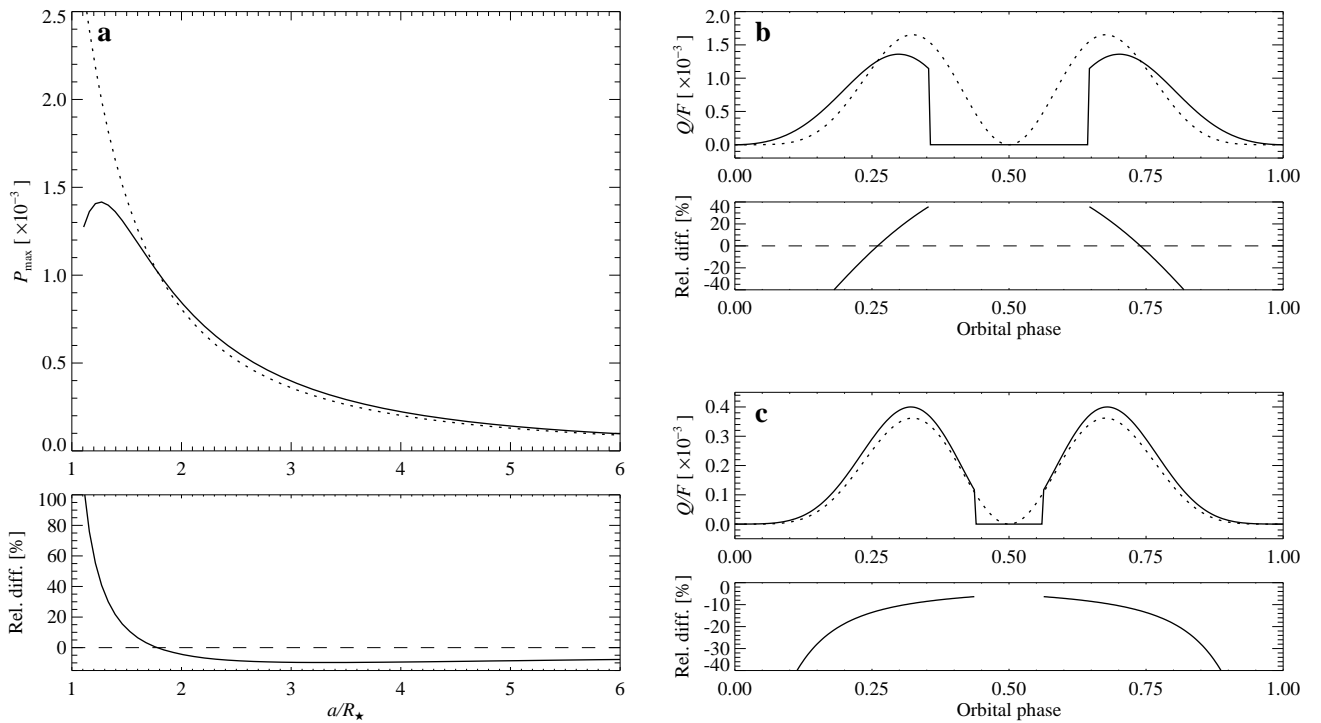}
  \caption{Comparison of models with extended (solid) and point-like (dotted) stars. {\bf a)} The panel illustrates the maximum degree polarization during one planetary orbit as a function of the semi-major axis $a$, which is given in units of the stellar radii. The relative difference is given in the panel underneath. {\bf b)} Stokes $Q$ curve at a distance $a/R_\star\!=\!1.4$ and relative difference between the extended (solid) and point-like (dotted) stellar models. Note that Stokes $U$ remains zero for the chosen parameters. {\bf c)} Same as panel {\bf b} but at a distance $a/R_\star\!=\!3.0$. The following parameters were chosen for the calculations: $i\!=\!90$\deg, $\Omega\!=\!90$\deg, $e\!=\!0$, $r\!=\!1$~$R_\mathrm{J}$, and, in the models with a finite sized star (solid), $R_\star\!=\!1$~$R_\mathrm{\odot}$. The dashed lines indicate the zero level.}
  \label{fig:starSize}
\end{figure*}

A first immediate effect of models with extended stars is the possible appearance of transits and secondary eclipses. During a secondary eclipse the planet moves behind the star as seen from the line-of-sight, so that Stokes $Q$ and $U$ drop to zero, assuming that the direct stellar light is unpolarized. This effect is illustrated in Fig.~\ref{fig:eclipses} where the inclination is varied. A secondary eclipse occurs if the inclination lies in the range $90$\deg$-i_\mathrm{e}\le i\le 90$\deg$+i_\mathrm{e}$, where $i_\mathrm{e}$ depends on the relative size of the star and the planetary orbit. The effect of the eclipse on the polarization curves can be significant and provides a very sensitive tool to diagnose the orbital inclination and the orbital phase. Note that the planetary transits lead to much smaller relative signatures in the intensity light curves because only the direct stellar flux is slightly reduced. In Fig.~\ref{fig:eclipses} the transits remain invisible although they were modeled in our calculations.

The direct comparison of the point-like and extended star models reveals additional differences of the curves apart from eclipses (Fig.~\ref{fig:starSize}). First we consider the maximum degree of polarization $P_\mathrm{max}$ reached during one orbit as a general measure to compare the two models (Fig.~\ref{fig:starSize}a). The most significant deviations naturally appear for small planetary orbits, while the two models converge for large orbits as expected. We find that the two curves representing the two models cross at about a semi-major axis $a/R_\star\!=\!1.8$. The exact location of this cross-over depends on the limb-darkening of the star, but always occurs around $a/R_\star\!=\!2$. With a solar limb-darkening the cross-over happens at $a/R_\star\!=\!2.3$.

For planetary orbits with semi-major orbits larger than $a/R_\star\!\approx\!2$ the point-like star model always somewhat underestimates the maximum polarization compared to the realistic extended star model. The difference slightly depends on the stellar limb-darkening, but amounts to about 10\% at the distance $a/R_\star\!\approx\!3$ and then continuously reduces for larger orbits to about 5\% at the distance of known hot Jupiters ($a/R_\star\!\approx\!10$) and below 1\% at $a/R_\star\!\approx\!60$ (almost 0.3~AU for solar analogs). These deviations arise because the solid angle occupied by the star (as seen from the planet) is properly calculated in the extended star model, which leads to a slightly increased incident flux at large distances.

For very small planetary orbits ($a/R_\star\!<\!2$) the point-like star model leads to a maximum polarization that is much too high, up to a factor of two higher than the realistic extended star model. At a first glance it might sound surprising that the extended star model results in a smaller maximum degree of polarization despite the planet being much closer to the stellar surface. At $a/R_\star\!=\!2$ the distance planet-stellar surface in the extended star model is halved compared to the point-like model, which in principles multiplies the incident flux by a factor of four. However, at these short distances only a part of the stellar hemisphere is visible from the planet, and the effective distance to points close to the limb is still relatively large, which almost compensates the proximity to the disk center, although the incident flux stays a bit stronger in the extended star model. 

The crucial point is that in the extended star model the incident radiation does not arrive anymore parallel to the planet, which results in contributions to both Stokes $Q$ and $U$ and partial cancellation of polarization, while in the point-like model all incident (parallel) rays add up to the same direction of linear polarization. Assuming that light from the disk center gives rise to positive Stokes $Q$ and zero $U$ for the distant observer, then light from the limb starts to contribute mainly to Stokes $U$ if the scattering plane is rotated by more than 22.5\deg, although the total Stokes $U$ cancels out due to symmetry reasons. If the planet orbits very closely to the stellar surface, some incident radiation would even arrive with angles greater than 45\deg\ relative to the stellar disk center, which would in turn lead to contributions to negative $Q$. At $a/R_\star\!=\!2$ the star has an angular size of about 25\deg\ viewed from the planet. This is just the limit above which contributions to positive Stokes $Q$ start to decline significantly. It is thus not surprising that the degree of polarization in the extended star model is actually smaller if the planet moves very closely to the stellar surface.

Let us now look at the differences of the shape of the polarization curves resulting from the point-like and extended star models, respectively (Fig.~\ref{fig:starSize}b\&c). For large orbits, i.e.\ beyond the cross-over distance of the two models, the maxima of the two light curves occur at about the same phase angles (Fig.~\ref{fig:starSize}c). The comparison of the maximum degree of polarization $P_\mathrm{max}$ shown in Fig.~\ref{fig:starSize}a is thus sufficient to estimate the deviations of the two models: the polarization curve is just scaled according to the relative difference of $P_\mathrm{max}$, except that we have to consider the possible secondary eclipses in the extended model. For very small orbits, i.e.\ inside the cross-over distance, the shape of the light curves start to differ more significantly between the two models (Fig.~\ref{fig:starSize}b). The maxima of the light curve start to shift toward maximum elongation in the extended model (solid curve). In general the position of the polarization maxima is determined by optimizing the effect of two factors, the increasing phase function (Eq.~(\ref{eq:phaseFunction})) and the optimal scattering angle leading to particularly large polarization, which is 90\deg\ for Rayleigh scattering. In the extended star model with very small orbits the planet is illuminated from different directions, giving contributions to the scattered light with various phase and scattering angles. Contributions from stellar surface elements that lie close to the observer gain particular importance because the corresponding phase function is largest compared to the rest of the stellar surface. This effectively shifts the polarization maxima towards smaller phase angles. The polarization curves can easily differ by 20--40\% (Fig.~\ref{fig:starSize}b). In fact we find that the comparison of $P_\mathrm{max}$ (Fig.~\ref{fig:starSize}a) underestimates the relative errors introduced by the point-like star model if $a/R_\star\!<\!2$.


\section{Fitting procedure to simulated data}\label{sec:fits}

To derive the orbital parameters from the polarization data of extrasolar planets we performed a $\chi^2$-minimization. With the multitude of different Stokes $Q$ and $U$ phase curves presented in Sect.~\ref{sec:lightcurves} one can wonder to what extent they are unique and how easily the correct parameters can be retrieved. In this section we test the robustness of the fitting procedure. We consider two cases of circular and elliptical orbits with simulated data and try to recover the original parameters with $\chi^2$-minimization.

The orbital phases of the simulated data were randomly chosen within 10 bins of the orbital period to mimic the situation with real observations in which one would try to obtain a full phase coverage without actually achieving identical time steps. For these phases and a set of selected orbital and planetary parameters we computed Stokes $Q/F$ and $U/F$, which defined our input or true polarization curves. Then we applied a random Gaussian noise to the true curves resulting in a set of simulated data with different signal-to-noise levels in the two studied cases. For a given signal-to-noise ratio we defined the signal to be the difference between the maximum and minimum of the curve either in Stokes $Q$ or $U$, whichever was greater, while the noise level corresponded to one standard deviation of the Gaussian distribution.

\begin{figure*}
  \centering
  \includegraphics[width=17cm]{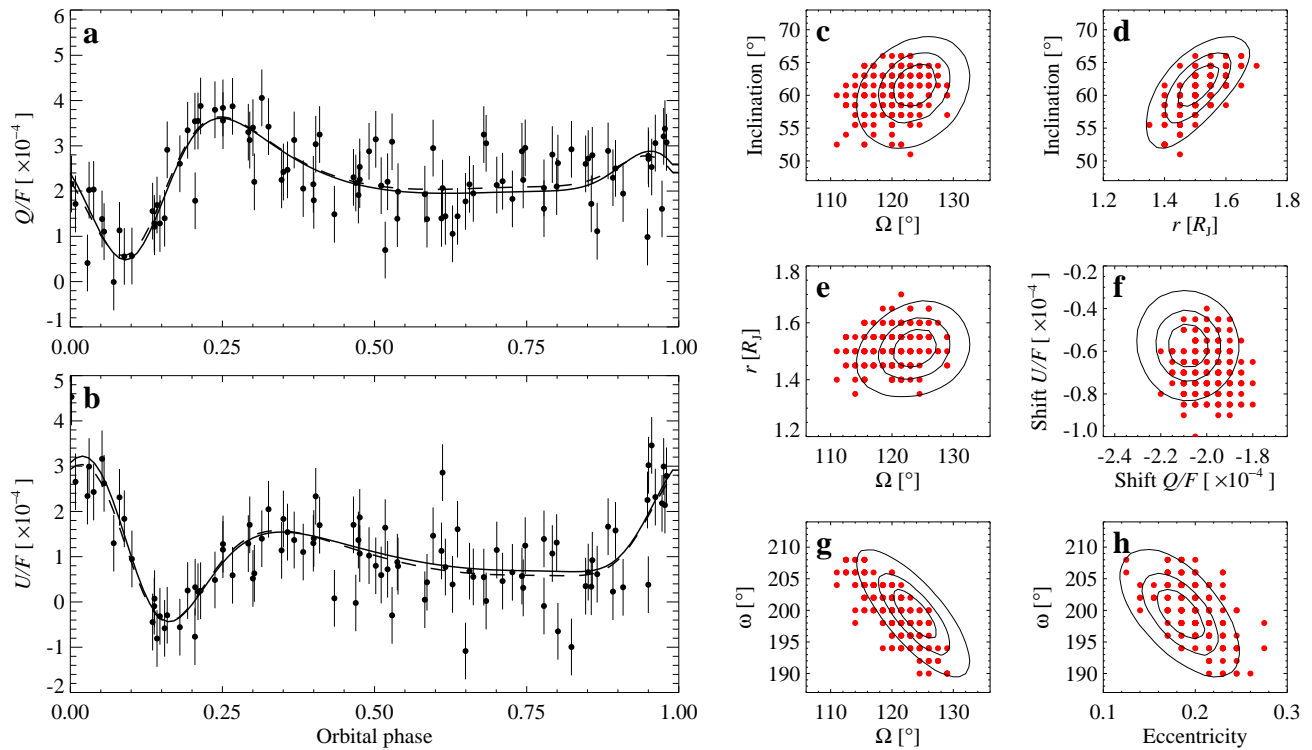}
  \caption{Results of the fitting procedure using simulated observations with $S/N\!=\!5$. {\bf a)} and {\bf b)} Input curves (solid) for Stokes $Q$ (top) and $U$ (bottom), from which the simulated data (dots) were generated by adding Gaussian noise with $S/N\!=\!5$, and the curves resulting from the best fit (dashed) to these data. Input and best-fit parameters are listed in Table~\ref{table:parameters} (case 1). The orbital phase zero is defined by the periastron passage. {\bf c)} to {\bf h)} $\chi^2$-contours (solid) as functions of the seven free parameters illustrated by a representative selection of two-dimensional projections within the parameter space. The contours show the differences to the minimum value for the levels $\Delta\chi^2\!=\!1.00$, 2.71, and 6.63, whose projections to the individual parameter axes correspond to the 68.3\%, 90.0\%, and 99.0\% confidence intervals, respectively. The Monte Carlo sample solutions are illustrated with dots. [See the electronic edition of the Journal for a color version of this figure.]}
  \label{fig:fit_sn5}
\end{figure*}

\begin{figure*}
  \centering
  \includegraphics[width=17cm]{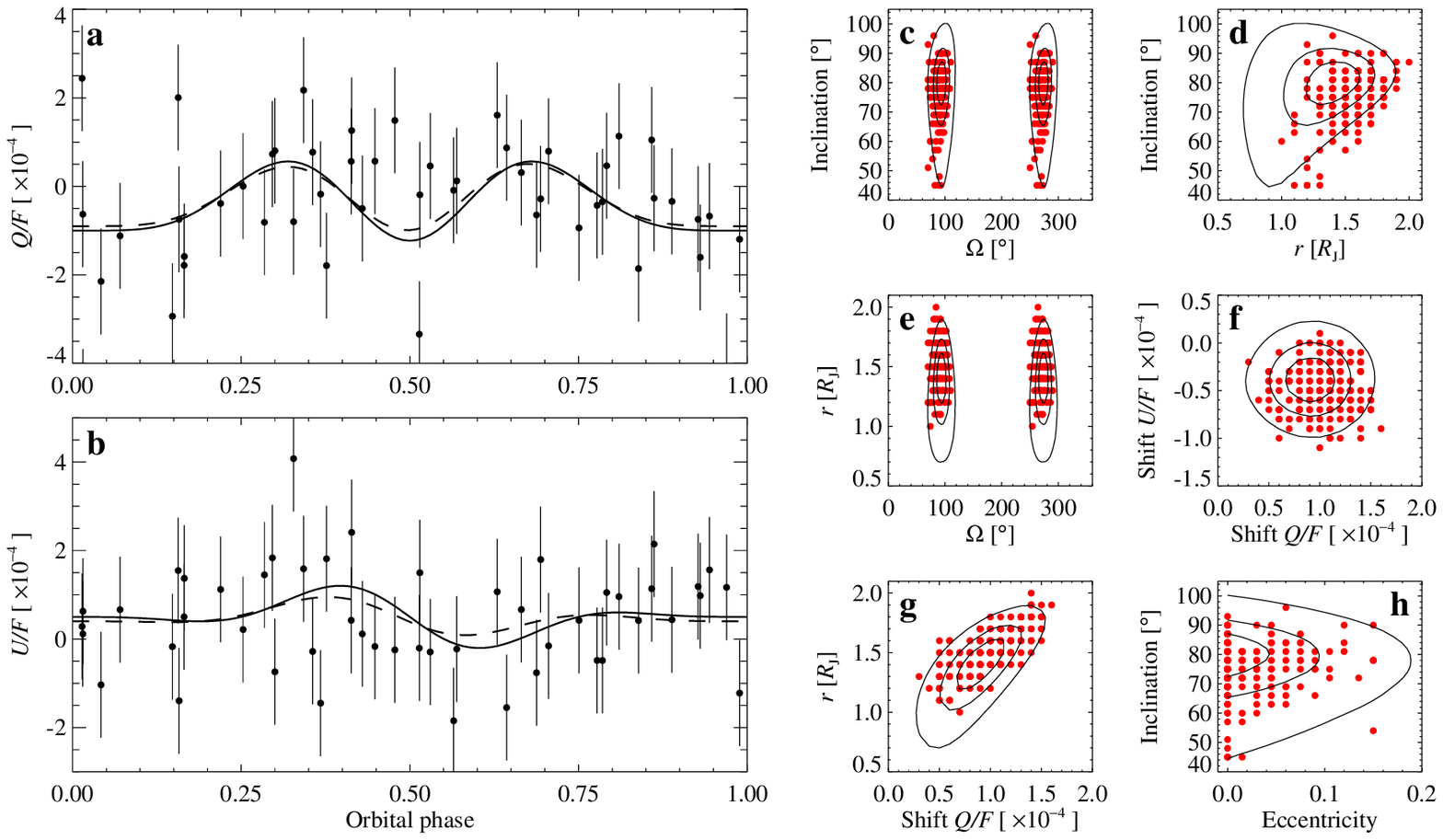}
  \caption{Same as Fig.~\ref{fig:fit_sn5}, but for different input parameters and $S/N\!=\!1.5$ applied to the simulated data. Input and best-fit parameters are listed in Table~\ref{table:parameters} (case 2). The orbital phase zero is given here by the greatest phase angle, since we have assumed a circular orbit. Note the ambiguity in $\Omega$ illustrated in panels c) and e).}
  \label{fig:fit_sn1.5}
\end{figure*}

Error bars of the best-fit parameters were evaluated with the help of Monte Carlo simulations. In the case of simulated data we have the advantage to know the ``true'' solution which is given by the chosen input parameters. For both considered cases we calculated 200 samples of measurements by applying Gaussian noise to this ``true'' solution in the same way as described above for the simulated data. In fact, the simulated observations (Figs.~\ref{fig:fit_sn5} and \ref{fig:fit_sn1.5}, panels a and b) can be considered as particular examples of such Monte Carlo samples. The $\chi^2$-minimization was applied to each Monte Carlo sample, resulting in 200 sets of best-fit parameters. The standard deviations from the true values define the 1$\sigma$ error bars for every free parameter.

\begin{table*}
\begin{minipage}[t]{\textwidth}
\caption{Parameters of the models used to test the fitting procedure with simulated data.} 
\label{table:parameters} 
\centering 
\renewcommand{\footnoterule}{}  
\newcommand\T{\rule{0pt}{2.2ex}}
\begin{tabular}{c c c c c c c c c c} 
\hline\hline 
Case & $S/N$ & $a$  & $r$  & $i$  & $\Omega$  & $e$ & $\omega$  & Shift $Q/F$  & Shift $U/F$  \\
 & & [AU]& [$R_\mathrm{J}$]&[\deg] & [\deg]& &[\deg] & [$\times 10^{-4}$]& [$\times 10^{-4}$]\\ 
\hline 
1 (Fig.~\ref{fig:fit_sn5}) & 
5.0 &
$\begin{array}{c} 0.03\footnote{Fixed parameter} \T \\ 0.03^a \end{array}$ & 
$\begin{array}{c} 1.5 \\ 1.5\pm0.1 \end{array}$ & 
$\begin{array}{c} 60  \\ 61\pm3 \end{array}$ & 
$\begin{array}{c} 120 \\ 124\pm4 \end{array}$ & 
$\begin{array}{c} 0.2 \\ 0.19\pm0.03 \end{array}$ & 
$\begin{array}{c} 200 \\ 199\pm4 \end{array}$ & 
$\begin{array}{c} -2.0\\ -2.1\pm0.1 \end{array}$ & 
$\begin{array}{c} -0.7\\ -0.6\pm0.1 \end{array}$
\\
\hline
2 (Fig.~\ref{fig:fit_sn1.5}) & 
1.5 &
$\begin{array}{c} 0.03^a \T \\ 0.03^a \end{array}$ & 
$\begin{array}{c} 1.5\\ 1.4\pm0.2 \end{array}$ & 
$\begin{array}{c} 75\\ 80\pm9 \end{array}$ & 
$\begin{array}{c} 90\\ 94\pm8 \end{array}$ & 
$\begin{array}{c} 0.0 \T \\ 0.0\footnote{The uncertainty of $e$ is 0.04} \end{array}$ & 
$\begin{array}{c} -\footnote{Undefined parameter; fixed to 90\deg\ in the fitting procedure for eccentric orbits} \T \\ - ^c \end{array}$ & 
$\begin{array}{c} 1.0\\ 0.9\pm0.3 \end{array}$ & 
$\begin{array}{c} -0.5\\ -0.4\pm0.3 \end{array}$
\\
\hline 
\end{tabular}
\end{minipage}
In each case the top row refers to the true parameters employed to simulate the data, while the bottom row gives the result obtained from the $\chi^2$-minimization. The 1$\sigma$ errors correspond to the standard deviation of each parameter as obtained from the Monte Carlo simulations.
\end{table*}

In the first simulated case (Fig.~\ref{fig:fit_sn5}) we assumed an elliptic orbit, given by 100 data points, and applied a moderate signal-to-noise ratio $S/N\!=\!5$ to the original curves (solid, panels a and b). In the fitting procedure we simultaneously optimized seven parameters: the planetary radius $r$, the orbital inclination $i$, the position angle of the ascending node $\Omega$, the eccentricity $e$, the longitude of the periastron $\omega$, zero point shifts of Stokes $Q$ and $U$. These shifts in the polarizations scales can account for background polarization either due to circumstellar dust or interstellar scattering, although the most ideal situation is of course to measure possible constant off-sets.

The Stokes $Q$ and $U$ phase curves are very well reproduced by the fitting procedure (Fig.~\ref{fig:fit_sn5}a\&b, dashed lines). All seven parameters returned by the $\chi^2$-minimization agree well with the original input parameters (Table~\ref{table:parameters}). In panels c--h of Fig.~\ref{fig:fit_sn5} we show two-dimensional projections of $\chi^2$ in the parameter space. The minimization procedure resulted in a minimum value $\chi^2_\mathrm{min}\!=\!1.03$ per degree of freedom. There exists however a second identical minimum, not indicated in the figure, in which $\Omega$ is shifted by 180\deg. Apart from this ambiguity, which is discussed in Sect.~\ref{subsec:OMEGA}, the minimum of $\chi^2$ is unique and well constrained. We found that the moderate signal-to-noise $S/N\!=\!5$ is already sufficient to determine the true parameters within a few percent.

The best-fit solutions of the Monte Carlo samples (dots in panels c to h, Fig.~\ref{fig:fit_sn5}) are of course centered around the true solution, which explains the small relative shift of the $\chi^2$-contours, which indicate the best-fit solutions for the simulated data. As statistically expected this shift is smaller than the 1$\sigma$ error.

In the second test case (Fig.~\ref{fig:fit_sn1.5}) we considered only 50, i.e.\ half as many, data points and we considerably reduced the signal-to-noise ratio down to 1.5, which is more representative of the data by \citet{berdyuginaetal2008a}. Here we assumed a circular orbit so that the longitude of the periastron becomes undefined. For the $\chi^2$-minimization we nonetheless kept the eccentricity $e$ as a free parameter, but we fixed $\omega$ to 90\deg. For very small eccentricities $\omega$ is not well constrained by the data but has also negligible influence on the polarization curves. Therefore, we are left with six free parameters.

Even under these more difficult circumstances the fitting procedure proved to be very robust. The original curves are again quite well reproduced (Fig.~\ref{fig:fit_sn1.5}a\&b) and the input parameters well identified (Table~\ref{table:parameters}). The $\chi^2$-minimum ($\chi^2_\mathrm{min}\!=\!1.19$ per degree of freedom) is still well defined but not as pronounced and broader than in the first test case, which increases the uncertainties in the fitted parameters. Note the ambiguity in $\Omega$, which is visible in panels c and e (Fig.~\ref{fig:fit_sn1.5}). We conclude that the data of the second test case are still good enough to accurately approximate the orbital parameters despite the small signal-to-noise.


\section{Conclusions}\label{sec:conclusions}

We developed and tested the tools for the interpretation of polarimetric data from extrasolar planets. We employed (i) an analytical approach based on the Lambert sphere approximation for intensity combined with Rayleigh scattering for polarization and (ii) a numerical method for self-consistent polarized radiative transfer in a scattering and absorbing atmosphere. We demonstrated that these two approaches lead to somewhat different intensity phase functions but negligibly different shapes of polarization curves.
We focused on the influence of various orbital parameters on the Stokes $Q$ and $U$ curves for the general case of elliptic planetary orbits and carried out a parameter study. The parameters include the inclination, longitude of the ascending node, eccentricity and the periastron longitude of the planetary orbit. The shapes of the polarization curves are not only very sensitive to changes in all of these parameters, but are also unique for every set of parameters except for a 180\deg\ ambiguity in $\Omega$ that results from the intrinsic properties of Stokes $Q$ and $U$.

We found that the finite size of the parent star should be correctly taken into account for very small orbits. Otherwise the approximation of a point-like star would lead to errors in modeling polarization exceeding 10\% if the planet resides closer than about three stellar radii from the surface of the star. For known hot Jupiters this error is reduced to about 5\%, which might be acceptable depending on the polarimetric precision of the data.

Fits to data with a $\chi^2$-minimization proved to be very stable even for data with a signal-to-noise ratio close to one if a sufficient number of data points (typically 50 or more) are available. The Stokes $Q$ and $U$ curves are sufficiently unique that the fitting procedure can reliably determine up to seven free parameters simultaneously.

Polarimetric studies of extrasolar planets open up new exciting possibilities to explore these remote worlds, even when considering merely orbital parameters. Our analysis can derive the inclination of the orbit for example, which improves the determination of the planetary masses, for which currently often only a lower limit $M\sin i$ is known. The Stokes $Q$ and $U$ curves also constrain $\Omega$, the position angle of the ascending node of the planetary orbit. The orbital parameters of the planet could also lead to better estimates of the orientation of the stellar spin axis, which appears to be well aligned with planetary orbits. In possible transiting systems of hot Jupiters orbiting giant stars secondary eclipses can lead to drops in polarization on the order of 50\% of the total signal, compared to a reduction in the flux of about 1\% during transits. The most significant advances in the study of extrasolar planets will result however from future observations with greater polarimetric sensitivity that will allow us to study directly the chemical and physical properties of the planetary atmospheres.


\begin{acknowledgements}
SVB acknowledges the EURYI (European Young Investigator) Award provided by the European Science Foundation (see www.esf.org/euryi) and SNF grant PE002-104552.
\end{acknowledgements}


\bibliographystyle{aa}
\bibliography{09970}


\end{document}